# TRANSACTION COSTS AND INSTITUTIONAL CHANGE OF TRADE LITIGATIONS IN BULGARIA

**Shteryo Nozharov**[∗]

**Petya Koralova – Nozharova**[∗]

**Abstract**: The methods of new institutional economics for identifying the transaction costs of trade litigations in Bulgaria are used in the current paper. For the needs of the research, an *indicative model*, measuring this type of costs on microeconomic level, is applied in the study. The main purpose of the model is to forecast the rational behavior of trade litigation parties in accordance with the transaction costs in the process of enforcing the execution of the signed commercial contract. The application of the model is related to the more accurate measurement of the transaction costs on microeconomic level, which fact could lead to better prediction and management of these costs in order market efficiency and economic growth to be achieved. In addition, it is made an attempt to be analysed the efficiency of the institutional change of the commercial justice system and the impact of the reform of the judicial system over the economic turnover. The augmentation or lack of reduction of the transaction costs in trade litigations would mean inefficiency of the reform of the judicial system.

---

[∗] Shteryo Nozharov, Ph.D., Chief Assistant professor in Department of Economics, University of National and World Economy, Bulgaria; e-mail: nozharov@unwe.bg
Dr. Nozharov is a Secretary of the Section of Economics-Union of Scientists in Bulgaria

[∗] Petya Koralova-Nozharova, Ph.D., Chief Assistant professor in Department of Economics and accountancy of transport, Todor Kableshkov Higher School of Transport, Bulgaria; e-mail: pkoralova@vtu.bg

# TRANSACTION COSTS AND INSTITUTIONAL CHANGE OF TRADE LITIGATIONS IN BULGARIA

Shteryo Nozharov[*]
Petya Koralova – Nozharova[**]

**Introduction**

The share of shadow economy in Bulgaria as a percentage of GDP for the period 1999-2007 is 35% and the country is in the first half of the ranking among 120 countries (Schneider, Buehn and Montenegro, 2010). In the period 2008-2014 the share of the shadow economy as a percentage of GDP in Bulgaria is almost the same -31% and the country take the first place in the ranking among other European countries and get ahead of many developing economies (Schneider, Raczkowski and Mróz, 2015). For example, the shadow economy as a percentage of GDP in Germany is 13.3%, and in Slovakia it is 14.6% for the same period.

The current study assumes that some of the reasons for the large share of shadow economy in Bulgaria as a percentage of GDP due to the large transaction costs level. In this regard, *the main objective of the research* is to identify and measure those transaction costs which are typical for the trade litigations in Bulgaria (Code of Civil Procedure 2007, art. 365-378) in order to be minimized. In accordance with the institutional economics theory, the trade litigation is defined as a type of transaction cost in relation to the process of enforced execution of the commercial contract (Dahlman, 1979). In this regard, the institutional failures of the court proceedings could increase the total transaction costs for both parties of the trade.

The limitations and the format of the research impose its subject to be narrowed to studying the transaction costs of trade litigations in Bulgaria, caused by the institutional failures of the forensic accounting expertise.

The main thesis of the study is that the failures of the institutional environment of trade litigations increase transaction costs and cause unpredictability of the commercial contracts performance. This leads also to reduced economic activity of the companies.

---
[*] Shteryo Nozharov, PhD., Chief Assist. Prof., Department of Economics, UNWE – Sofia, Dr. Nozharov is a Secretary of the Section of Economics-Union of Scientists in Bulgaria. email: nozharov@unwe.bg
[**] Petya Koralova-Nozharova, PhD., Chief Assist. Prof., Department of Economics and Accountancy of Transport, Todor Kableshkov Higher School of Transport – Sofia, email: pkoralova@vtu.bg

In accordance with the thesis, the following tasks will be solved:
1. The failures of the institutional environment of trade litigations in Bulgaria will be identified;
2. An econometric model for transaction costs measurement of trade litigations as a result of the institutional environment failures will be elaborated.

The significance of the research is related to the development of the transaction costs theory. There are many studies which examine transaction costs in the first two phases of the commercial contract development: the phase of finding counterparty and the phase of negotiation and signing of the contract. However, there are few studies related to the third phase of the contract, which examines the transaction costs caused by its execution. Moreover, the studies, related to the transaction costs caused by the enforced commercial contract execution by the Court are fewer. In the literature review this fact will be explained.

Having in mind the aforementioned, the significance of the research is related both to the transaction costs identification as a result of the enforced execution of the commercial contract and the implementation of an econometric model for their measurement. This will help the total transaction costs of trade litigations to be measured more accurately.

**Literature Review**

There are some papers that identify the concept of the transaction costs in the court proceedings. For example the research of Kessler and Rubinfeld (2007) where through an empirical analysis of the court records, the impact of the law (implication and alternative amendments) over the trade litigation and the related transaction costs is examined. There it is made a comparison among the transaction costs of the formal litigation and the opportunity for out-of-court dispute resolution. Transaction costs are identified in terms of benefits for both parties in the transaction as a result of an out-of-court dispute resolution. The model itself consists of those types of transaction costs which are not related to the risk of wrong court decision. The model does not identify the institutional environment failures and the ways they could be overcome.

Another similar publication is that of Reda (2011). He finds out that trade litigation could redirect the transaction costs (change their places) for both parties. Moreover, the transaction costs related to bringing claims against large corporations are much greater. This fact impacts not over the justice but also the economic turnover. He also examines the impact over the costs and the outcome of the legal case when the legal representation is done by large law companies or self-employed lawyers. The main principles for identifying the trade litigation as an effective one: fairness, speed litigation and accessibility, are also studied in this publication. The main accent of the study is put on the reduction of the court costs as a result of the improvement of the access to justice. The cited author has not presented a statistical or econometrical model, related to the court costs measurement, in his publication. The author examined various elements of the court costs and he put them into various categories. He referred to statistical data and other authors' publications. The analysis of the transaction costs takes small part of Reda's research.

This issue is also addressed in OECD publications (e.g. Palumbo et al. 2013). There it is found that the case length could be an obstacle and hamper the economic activity. It is also stated that the high length of overall cases results in higher transaction costs for companies. The structure of the litigation costs, as well as the courts' structure is used as main determinants for the needs of the cited publication. The predictability of court decisions, which reduces both the case length and likelihood of appeals, is introduced as an additional determinant. In the conclusions, it is stated that the well-functioning judicial systems are key factor for the economic development. They reduce the transaction costs and guarantee security for the contracts' execution as well as reduce the opportunistic behavior of the economic agents. In the cited publication, transaction costs are examined on macroeconomic level and the analysis is made on the basis of a comparative analysis among different countries.

Among the Bulgarian authors, there could be mentioned the publications of Sedlarski (2008) and of Tchobanov, Egbert and Sedlarski (2008). In both publications, the justice is defined as a transaction service, provided by the public sector. The authors find that the delayed reform of the judicial system leads to obstacles for the courts to enforce the law, which fact impacts over the sum of transaction costs of litigations and the economic activity. The transaction costs are studied on macroeconomic level and a relationship between the dynamics of the transaction costs and economic growth is made.

The main purpose of the current research is transaction costs to be examined on microeconomic level.

## 1. Identifying the failures of the institutional environment

In accordance with the limitations of the current research, the identification of the transaction costs will be focused on the institutional system of the forensic accounting expertise of trade litigations. Moreover, the institutional system will be studied in accordance with the last amendments of the Bulgarian legislation since 2017.

The failures of the institutional environment are summarized as a result of the analysis of commercial law cases which are publicly available in the legal-information systems (e.g. APIS, SIELA and etc.).

**First institutional failure: lack of competence of the forensic accounting expert witnesses**

This institutional failure is related to the opportunity an expert witness with a three-year college education (currently referred to the educational degree "professional bachelor") and with a vague practical experience in the field of Accounting to check the work of an experienced certified accountant or of an expert with a chartered certified accountant qualification (ACCA). When analyzing the commercial court cases, the authors have identified similar cases in the Official Journal of the Republic of Bulgaria concerning expert witnesses.

In accordance with the most recent amendments of Ordinance 2/2015 regulating the statutes of the expert witnesses, every expert witness must hold an educational degree or qualification concerning the type of expertise he works on and have at least 5 years' experience in the relevant specialty (Ordinance 2/2015, **art.7, para.1-2**). Ordinance 2/2015 is too general and does not distinguish individual requirements for carrying out different types of expertise, including forensic accounting expertise. Consequently, the general requirements for financial statements compilation in accordance with the Accountancy Act must be applied, which means that these requirements are the minimal requirements for an expert to be nominated as a forensic accounting expert witness (Accountancy Act 2015, **art.18, para.1**). The most recent amendments of the Accountancy Act, concerning the education of financial statement compilers assume that persons holding college economic education or higher economic education could be hired as financial statement compilers. In this regard, the minimum qualification requirements for a forensic accounting expert witness, who will check the work of experts holding master's degree in accounting or are experienced certified accountants, regards only to having a college economic education or just graduated any economic specialty no matter it is in the field of accountancy.

In accordance with the cited law acts, together with the minimum requirements for education of the expert witnesses, there is also a requirement for 5 years' experience in the relevant specialty. As in the Ordinance 2/2015, it is not stated what kind of experience is defined as an experience in the field of accountancy, consequently there will be applied the general rules of the Accountancy Act, concerning the requirements for financial statement compilers. In accordance with the Accountancy Act as an experience in the field of accountancy is assumed the experience in the field of internal and external audit, financial inspection, tax audits as well as experience as a teacher in accountancy and financial control. However, in the Accountancy Act it is not specified which occupations are assumed as an experience in the field of accountancy. These occupations are defined in the National Classificatory of the Professions and Occupations in the Republic of Bulgaria (2011). According to this act except for accountants, other occupations in the field of accountancy which are assumed as an experience in the field of accountancy are also cashiers and invoice clerks.

In this regard, the minimum requirements for an experience of forensic accounting expert witnesses are formal. Actually, the Ordinance 2/2015 does not contain any specific requirements for forensic accounting expertise and also does not refer to any other applicable legal act. Even though one of the requirements of the Ordinance for taking an occupation in the field of accountancy is to be submitted a copy of the employment record book of the applicant, it is still unclear what kind of experience is assumed as an experience in the field of accountancy. It is not also clear if people who certify the registration of expert witnesses use the requirements of other legal acts as analogy, such as those concerning the registration of the financial statements compilers or the National Classificatory of the Professions and Occupations in the Republic of Bulgaria.

As a result of the aforementioned, it is proved that the legislation in the Republic of Bulgaria assumes registration of forensic accounting expert witnesses who have college economic education or high education in the field of accountancy and an experience in occupations such as cashiers or invoice clerks. These persons are assigned by the Commercial Court to check the accounting audit of large corporations and consortiums whose financial reports are prepared by experienced certified accountants or by experts with a chartered certified accountant qualification.

**Second institutional failure: the possibility of a conflict of interests and non-objectivity in the assignment of a second expertise by another forensic accounting expert witness.**

This is the case when the initial expert witness who has prepared the forensic accounting expertise and the newly appointed expert witness who is accustomed to repeat the expertise are members of the same joint venture, established for private benefit. There are several alliances where most of the expert witnesses in the Republic of Bulgaria belong to. These organizations are officially registered in the Court as the content of their statutes is available in the relevant Court registers. For example on the web page of one of the expert witnesses organizations the following text is written: „the organization is *established on the basis of an existing structure of professional relationships among acting, recognized and proved experts, expert witnesses and experts in the field of court arbitration.* "

In the statute of the organization it is clearly written that all of its members are in a relationship one to another. In addition, the organization is established for private benefit by carrying out an economic activity which results in property interests for its members:

*„…The alliance of the expert witnesses ...*is specialized in the provision of *highly professional expert services and consultations* by a team of expert witnesses *and experts in the relevant fields"*, *„The expert witnesses .... work as a team, which fact guarantees the high quality of the services provided "*.

In general, the publicly provision of services by these organizations make them economic entities, moreover they are registered in the Court as organizations for private benefit. It is also written in their statutes that the *expert witnesses work as a team* in the provision of services. Once such organisations are economic entities, even though they do not share profit, the money, earned through the economic activity of the organisations could be spent for things that are of benefit for their members. For example, the members of such organizations could benefit from attending education courses for professional qualification, expensive equipment, buildings in different cities which make easier the work of the forensic accounting expert witnesses and etc. At the same time these experts receive direct payments by the Court for the work they have done.

Having in mind the aforementioned, it is obvious that the expert witnesses who has to repeat the initial forensic accounting expertise, done by his colleague who is a member of the same alliance of the expert witnesses, will have an interest to confirm the initial expertise. Otherwise, this expert witness will risk to be deprived from using the expensive equipment, or attending qualification courses as a member of the alliance. This institutional failure could be overcome if the Court of Justice constitutes the membership in alliances of expert witnesses which are established for private benefit, as a potential conflict of interests.

## 2. Definition of the research model

The research model is developed on the basis of the model presented by Kessler and Rubinfeld (2007) which was mentioned in the introduction of the current study. The cited model compares the benefits of the likelihood the civil claim to be won and the benefits of signing a court settlement agreement:

$$G = (Tp - Td) + (c_{tp} + c_{td}) > 0 \quad (1)$$

where

**Tp (Td)** = subjective value to plaintiff (defendant) associated with a successful trial outcome; $c_{tp}$ $(c_{td})$ = cost to plaintiff if the case is settled

Then according to equation (1) the case will be settled if G>0.

The model of Kessler and Rubinfeld (2007) does not take into account the risk coefficient of wrong court decision. On the other hand, the model takes into account the sum of the subjective value to plaintiff (defendant) and the costs if the case is settled. The authors of the current study think that the concept of Kessler and Rubinfeld's model should be essentially modified. The court costs must be deducted from the net subjective value to the plaintiff (defendant) and at the same time the risk of wrong court decision must also be taken into account.

As a result of equation (1), it could be created a model for measuring the transaction costs in the following conditions: the execution of the signed contract is enforced by the Court, the costs are discussed before the beginning of the trial and the contracting parties have made a strategic and rational choice what to do. The model considers the viewpoint of the plaintiff. (This is the party who considers its duties in the contract to be discharged). For that purpose, the following equation could be used:

$$TC = [(Tp-Td)-(c_{tp1} + c_{td1})] \times Cf_r \quad (2)$$

Where:

$c_{tp1}$ $(c_{td1})$ = direct trial costs to the plaintiff (defendant),

$$Cf_r \text{ (risk coefficient)} = (Z + Kb + t_l) - (Y + Ka + t_s) \quad (3)$$

**Z** is the likelihood the forensic accounting expertise to be incompetent or dishonest;

**Kb** is the absence of predictability and equation of results of the Court decision in accordance with art. 365-378 of the Civil Code of Procedure;

$t_l$ is the long duration of trial (over one year), reported by the commercial law cases;

**Y** is the presence of preliminary court precautionary measures in accordance with art. 390 of the Code of Civil Procedure;

**Ka** is the presence of predictability and equation of the results of the Court decisions reported by the commercial law cases in art. 365-378 of the Code of Civil Procedure;

$t_s$ is the short duration of trial (under one year), reported by the commercial law cases;

The value of each determinant in the equation for measuring the risk coefficient (**Cf$_r$**) equals 1 in the presence of the condition or equals 0 in the absence of the condition.

**Empirical testing**

*First hypothesis*: The plaintiff has brought a claim for 100 000 euro and the defendant confirm the claim or expect the Court to confirm only 80% of the claim. The trial costs are 9 000 euro for each party in the contract. The values of the determinants necessary for measuring of the risk coefficient are: $Z=1$, $Kb=1$, $t_l=1$; $Y=1$, $Ka=0$, $t_s=0$; After the necessary calculations are done, the transaction costs for settling the agreement before the beginning of the trial according to this hypothesis are expected to be 4 000 euro, which sum is 4% of the sum of the claim. This means that the transaction costs of the process of enforced execution of the contract are low. In this regard the rational behavior of the plaintiff is to bring the lawsuit but the rational behavior of the defendant is to suggest a settlement agreement.

*Second hypothesis*: The plaintiff has brought a claim for 100 000 euro and the defendant confirm the claim or expect the Court to confirm only 50% of the claim. The trial costs are again 9 000 euro for each party in the contract. The values of the determinants necessary for measuring of the risk coefficient are the same as in the first hypothesis: $Z=1$, $Kb=1$, $t_l=1$; $Y=1$, $Ka=0$, $t_s=0$; After the necessary calculations are done, the transaction costs for settling the agreement before the beginning of the trial according to this hypothesis are expected to be 64 000 euro, which is 64% of the sum of the claim. The transaction costs of the plaintiff are high and it is better for him to suggest the defendant a settlement agreement.

*Third hypothesis*: The plaintiff has brought a claim for 100 000 euro and the defendant confirm the claim or expect the Court to confirm only 90% of the claim. The authors of the current research consider this hypothesis as an impossible one. If the defendant confirms over 80% of the claim, it is rationally for him to suggest a settlement agreement to the plaintiff.

**Conclusion**

The current research identified the institutional failures of the legislation concerning trade litigations through analysis of the rules according to which forensic accounting expertise is made. This expertise is of great importance for the final court decision, because through an accounting procedure it is examined the claim of the party who pretends for incomplete and inaccurate commercial contract execution. As the Court is not specialized in the field of accountancy, the conclusions of the forensic accounting expert witnesses are of great significance for the court decisions. That is why the current research is focused on this issue. The proposed model shows that the overcoming of the institutional failures in the field of forensic accounting expertise will result in lower transaction costs as a result of the enforced execution of the commercial contract. As a result of the limitations of the study, many institutional failures of trade litigations were not mentioned. For example, is it allowed by the legislation a forensic accounting expert witness to make thousands of expertise per year and still there are no legal restrictions in this field and etc.

After the fulfilment of the institutional changes in the period 2015-2017, no improvement of the rules, regulating the settlement of trade litigations in Bulgaria is observed. In 2015 a new Ordinance, regulating the statute of forensic accounting expert witnesses and new Accountancy Act were adopted. Over the past two years, since the implementation of these legal acts, many failures in the field of forensic accounting expertise of trade litigations are identified.

Having in mind the aforementioned, the present research could raise a discussion about the opportunities how the transaction costs in trials to be reduced as well as how the effectiveness of the reform of the judicial system, measured through the institutional environment failures to be assessed.

On the next place, a model for measuring transaction costs of trade litigations is presented in the current study. This model could be successfully used for improving the concept of measuring the aggregated transaction costs. The risk coefficient was not presented as a component of the equations in other models, examined in the literature review. Consequently, the introduction of this risk coefficient in the presented model in the current research is the main contribution of the study. The empirical testing of the model shows that it could be successfully put into the practice.

There are many studies related to the market investigation and signing of contracts, but there are few studies related to the enforced contracts execution. The transaction costs related to the enforced contract execution (court costs) are still not studied in depth. Detailed statistical or econometrical studies are also missing in this field. Consequently, the accurate and overall identification of the transaction costs is hindered. The current research contributes to the accurate identification and measurement of the transaction costs of trade litigations and allows better empirical testing in the field of enforced contracts execution.

In this regard the present study is addressed to a wide range of experts. For example, in economic theory context it will be useful for the experts who carry out studies in the field of shadow economy, transaction costs, behavioral economics and microeconomics. The results of the present study could also be helpful to the practice, especially for the contract management, the reduction of hidden costs and to the risk management of contracts non-performance.